\documentclass[aps,showpacs,twocolumn]{revtex4}
\usepackage{amsbsy}
\usepackage{bbold}
\usepackage{graphicx,color}
\usepackage{amsfonts}
\usepackage{amsmath}
\usepackage{amssymb}
\usepackage{amsthm,bm}
\usepackage{dsfont}

\newcommand{\nn}{\nonumber}

\begin{document}
\title{Two components relativistic quantum wave equation for scalar bosons}
\author{Roland Combescot$^{a}$}

\affiliation{$^{a}$
Laboratoire de Physique, Ecole Normale Sup\'erieure, PSL Universit\'e, 
Sorbonne Universit\'e, Paris Cit\'e Universit\'e, CNRS, 24 rue Lhomond, F-75005 Paris, France}

\date{Received \today}

\begin{abstract}
We show that, in the relativistic regime, scalar bosons satisfy a quantum wave equation 
which is quite analogous to the Dirac equation. In contrast with the Klein-Gordon equation
it is first order with respect to time derivation. It leads in a regular way to the standard Schr\"odinger
equation in the non-relativistic limit. There are two components for the wave function in this representation for the scalar
boson, in a way completely analogous to the four components for the spin  $1/2$ fermion in the Dirac equation.
\end{abstract}
\maketitle

\section{Introduction}\label{intr}

As found in standard textbooks \cite{ll,bis}, one does not have for scalar bosons a satisfactory situation with respect to the generalization
to the relativistic regime of the standard Schr\"odinger wave equation $i\partial_t \psi = -(1/2m)\partial_{{\bf }r}^2 \psi$ (we set $\hbar=1$). 
The Klein-Gordon equation $-\partial^2_t \psi+\partial_{{\bf }r}^2 \psi=m^2 \psi$ (we set $c=1$) provides an appropiate relativistic
generalization, but it is second order with respect to time derivation. This is in contrast with the case of the spin $1/2$ fermion, where
the Dirac equation $\gamma^{\mu }_{ik}p_{\mu }\psi_{k}=m\psi_{i}$ (with $p_{\mu }=\{i\partial_{t},i\partial_{{\bf r}} \})$ provides the appropriate
relativistic equation, first order with respect to time derivation, corresponding to the Klein-Gordon equation. The price is naturally that
the wave function is actually a four-component object $\psi \equiv \psi_{i}$, corresponding to the dimensions of the four Dirac matrices 
$\gamma^{\mu }$. These four-components have a natural physical interpretation since one needs two components to describe the 
spin $1/2$, and in addition one has also two components to describe the positive and negative frequency eigenstates, which are interpreted physically in terms of particles and antiparticles. When one goes toward the non-relativistic regime, in the standard representation
the positive frequency components become large components, while the negative frequency ones become small components, leading for example to relativistic corrections.

In addition the Dirac equation leads to a conserved current $j^{\mu}$, with a temporal component $j^{0}= \sum_{i} |\psi_{i}|^{2}$ which
is positive. In the non-relativistic regime one recovers the fact that $|\psi|^{2}$ can be interpreted as the local presence probability of
the particle, an essential ingredient in the physical interpretation of the Schr\"odinger equation. In contrast, for scalar bosons, the
Klein-Gordon leads to a conserved current with temporal component $j^{0}=i(\psi^{*}\partial_{t}\psi-(\partial_{t}\psi^{*})\psi)$ which
is not in general positive, and so it can not be interpreted in general as a local density probability. As a result one does not have
an appropriate way to obtain the Schr\"odinger equation as the low momentum limit of a general relativistic formulation. This can
be considered as a manifestation of incompleteness for relativistic quantum mechanics \cite{niko}. In practice this state of affairs 
is not so much a problem since, in high energy physics, existing scalar bosons are short-lived so one does not actually face this problem.
Moreover relativistic wave equations are not so much used, and relevant problems are rather handled with quantum field theory.
Nevertheless this situation is clearly unsatisfactory, all the more since there are common stable scalar bosons such as atomic
$^4$He and, although they are usually in a low velocity regime where relativistic effects are unimportant, the standard 
Schr\"odinger equation which is used to describe their quantum properties can not be obtained as the low energy limit of a
relativistic quantum wave equation.

In this letter we point out that, in contrast with this common belief, there is for these scalar bosons a relativistic quantum wave
formulation which is quite analogous to the Dirac equation, and which leads in a regular way to the standard Schr\"odinger
equation in the non-relativistic limit. This formulation stems out from the known Kemmer representation \cite{kemm}, but this consequence is apparently overlooked. There are two components for the wave function in this representation for the scalar
boson, in a way completely analogous to the four components for the spin  $1/2$ fermion in the Dirac equation.

\section{Formalism}\label{form}

Let us start by recalling the Kemmer representation \cite{kemm,inter}. For a particle with mass $m$, the wave equation reads, 
quite similarly to the Dirac equation
\begin{eqnarray}\label{eqwave}
\beta_{\mu} \partial_{\mu} \psi + m \psi = 0
\end{eqnarray}
where the four $\beta_{\mu}$ matrices satisfy the following commutation rules, obtained by Duffin \cite{duff} from an analysis
of Proca equations \cite{proca},
\begin{eqnarray}\label{eqcommut}
\beta_{\mu}\beta_{\nu}\beta_{\rho} +\beta_{\rho}\beta_{\nu}\beta_{\mu} = \beta_{\mu} \,\delta_{\nu \rho} + \beta_{\rho} \,\delta_{\nu \mu}
\end{eqnarray}
where $\delta_{\nu \rho}$ and $\delta_{\nu \mu}$ are Kronecker symbols. For simplicity we keep Kemmer's pseudo-euclidian metric
$x_{\mu}=x,y,z$ for $\mu =1,2,3$, and $x_{4}=it$. Multiplying Eq.(\ref{eqwave}) by $\beta_{\rho}\beta_{\nu}\partial_{\rho}$
and making use of Eq.(\ref{eqcommut}), one finds the following relation
\begin{eqnarray}\label{eqrelatdpsi}
\partial_{\nu} \psi=\beta_{\mu} \beta_{\nu }\partial_{\mu} \psi
\end{eqnarray}
When $\partial_{\nu}$ is applied to this equation, and use is again made of Eq.(\ref{eqwave}), one finds that $\psi$ satisfies the Klein-Gordon
equation
\begin{eqnarray}\label{eqklgord}
\partial^{2}_{\nu}\,\psi=m^2 \,\psi
\end{eqnarray}
as expected from a fully relativistic formulation. This relativistic invariance of Eq.(\ref{eqwave}) is proved explicitly by Kemmer \cite{kemm}.

Let us come to the number of components of the wave function $\psi$, that we have not yet mentioned. In a way similar to the Dirac 
equation, one has to find possible irreducible representations of the algebra corresponding to the commutation relations 
Eq.(\ref{eqcommut}). In the case of the Dirac equation, the anticommutation relations of the Dirac matrices lead to the well-known four-dimensional representation. In the present case the result is somewhat more complicated \cite{kemm}. Except for a trivial one-dimensional 
representation, which is of no physical interest, there are only two different representations. One of them is 10-dimensional, and
describes a spin 1 particle. The corresponding equations Eq.(\ref{eqwave}) are just Proca equations, and the 10 components of the
wave function correspond to the four components of a 4-vector, together with the six components of the antisymmetric tensor obtained
from the derivatives of the 4-vector. In the special case where the mass of this spin 1 particle is zero, Proca equations reduce to
Maxwell equations, the 4-vector is the standard 4-potential $A^{\mu}$ and the tensor is the standard field tensor $F^{\mu \nu}$.

The other representation is the one which is of interest for us. It is 5-dimensional and it describes a spin zero particle. The five 
components correspond physically to the standard scalar field used to describe such a spin zero particle, together with its four 
spatio-temporal derivatives. The link between these components is expressed by Eq.(\ref{eqrelatdpsi}), which results from 
Eq.(\ref{eqwave}). Naturally this formalism is fully general, and we may choose in principle to make any unitary transformation we like.
However, for our purpose, it is more convenient to make use of the simple representation found by Kemmer \cite{kemm}.
In this representation the $\beta_{\mu}$ are hermitian matrices with matrix elements given by
\begin{eqnarray}\label{eqbeta}
\hspace{-1cm}(\beta_{k})_{ij}&=& \delta_{i,k+1}\,\delta_{j,5}+\delta_{j,k+1}\,\delta_{i,5} \hspace{10mm} k=1,2,3 \\ \nn
\hspace{-1cm}(\beta_{4})_{ij}&=&i \,(\delta_{i,5}\,\delta_{j,1}-\delta_{j,5}\,\delta_{i,1})
\end{eqnarray}
With these explicit expressions Eq.(\ref{eqwave}) becomes
\begin{eqnarray}\label{eqwavexpl}
\hspace{-1cm}-\partial_{t}\psi_{5}+m\psi_{1}&=&0  \\  \nn
\hspace{-1cm}\partial_{k}\psi_{5}+m\psi_{k+1}&=&0   \hspace{15mm} k=1,2,3 \\ \nn
\hspace{-1cm}\partial_{k}\psi_{k+1}+\partial_{t}\psi_{1}+m\psi_{5}&=&0
\end{eqnarray}
which displays the expected relations \cite{ll,bis} between the scalar $\psi_{5}$ and its spatio-temporal derivatives 
$\partial_{\mu}\psi_{5}$, the last equation leading to the Klein-Gordon equation.

\section{Two components equations}\label{2comp}

We look now more specifically to the equations which are first order with respect to time derivation, which is our initial purpose. We see that
only the first and the last equation in Eq.(\ref{eqwavexpl}) contain such a derivation. The remaining ones contain only space derivatives.
The resulting expression for $\psi_{k+1}$ can be carried into the last equation, so we are left with only two equations for $\psi_{1}$ and 
$\psi_{5}$. We write these equations in a more symmetrical way by introducing the combinations $\psi_{+}=(\psi_{5}+i\psi_{1})/\sqrt{2}$ 
and $\psi_{-}=(\psi_{5}-i\psi_{1})/\sqrt{2}$. This leads to
\begin{eqnarray}\label{eqwavexplpm}
\hspace{-1cm}i\partial_{t}\psi_{+}&=&m\psi_{+}-\frac{1}{2m} \partial^2_{k}(\psi_{+}+\psi_{-}) \\  \nn
\hspace{-1cm}i\partial_{t}\psi_{-}&=&-m\psi_{-}+\frac{1}{2m} \partial^2_{k}(\psi_{+}+\psi_{-})
\end{eqnarray}

For a particle at rest these equations reduce to $i\partial_{t}\psi_{\pm}=\pm m\psi_{\pm}$, giving $\psi_{\pm}=e^{\mp imt}$.
Hence $\psi_{+}$ and $\psi_{-}$ describe particles with respectively positive and negative energies. If we then consider a
particle with a positive energy and a small momentum, the time dependence of $\psi_{\pm}$ is dominantly $e^{-imt}$, so
we set as usual $\psi_{\pm}=e^{-imt}\,\varphi_{\pm}$ where the time dependence of $\varphi_{\pm}$ is weak compared to
$e^{-imt}$. From Eq.(\ref{eqwavexplpm}) this gives
\begin{eqnarray}\label{eqwavexplpma}
\hspace{-1cm}i\partial_{t}\varphi_{+}&=&-\frac{1}{2m} \partial^2_{k}(\varphi_{+}+\varphi_{-}) \\  \nn
\hspace{-1cm}i\partial_{t}\varphi_{-}&=&-2m\varphi_{-}+\frac{1}{2m} \partial^2_{k}(\varphi_{+}+\varphi_{-})
\end{eqnarray}
In the second equation the term $\partial_{t}\varphi_{-}$ is negligible compared to $2m\varphi_{-}$, and we have
$\varphi_{-} \simeq \partial^2_{k}(\varphi_{+}+\varphi_{-})/4m^2$. So in this regime $\varphi_{-}$ is small compared
to $\varphi_{+}$. Hence we have in this regime a large component $\varphi_{+}$ and a small component $\varphi_{-}$,
in full analogy with the situation of the Dirac equation in the non-relativistic regime. At the lowest order we may completely
neglect $\varphi_{-}$ compared to $\varphi_{+}$ in the first equation Eq.(\ref{eqwavexplpma}), and we naturally recover the
Schr\"odinger equation $i\partial_{t}\varphi_{+}=-\partial^2_{k}\varphi_{+}/2m$. If instead we keep in this first equation
the lowest order expression $\varphi_{-}= \partial^2_{k}\varphi_{+}/4m^2$ for $\varphi_{-}$, we obtain
$i\partial_{t}\varphi_{+}=-\partial^2_{k}\varphi_{+}/2m-\partial^4_{k}\varphi_{+}/8m^3$, which gives the expected first order
relativistic correction to the Schr\"odinger equation for $\varphi_{+}$. 
 
We can extend this handling to eliminate $\varphi_{-}$ and obtain an all orders expansion in powers of $1/m$, providing the
whole relativistic contribution to the equation for $\varphi_{+}$. 
Since the operators we deal with are commuting, we may handle them as scalars. It is then
convenient to use the short-hand notations $i\partial_{t} \to \epsilon $ and $-i\partial_{k} \to p$. Multiplying the second equation
by $p^2$ gives $(\epsilon +2m+p^2/2m)\,p^2 \varphi_{-}=-p^4 \varphi_{+}/2m$. This can be carried into the first equation, 
multiplied by $(\epsilon +2m+p^2/2m)$, to eliminate $\varphi_{-}$. This leads to 
$(\epsilon ^2 +2m\epsilon)\varphi_{+}=p^2\,\varphi_{+}$. Comparing with the expansion of $(1+p^2/m^2)^{1/2}$ in powers of
$1/m$ gives us the power expansion of $\epsilon $ which is solution of this equation. Coming back to the initial notations this
leads to the full expansion \cite{expansion}
\begin{eqnarray}\label{eqwavegen}
i\partial_{t}\varphi_{+}=-m\, \sum_{n=1}^{\infty}\frac{1}{2n-1}\,\frac{(2n)!}{(n!)^2}\,\Big(\frac{\partial^2_{k}}{4m^2}\Big)^n\,\varphi_{+}
\end{eqnarray}
 
Coming to the physical interpretation, we have now to consider the conserved current. From Kemmer formalism one finds that
its temporal component is given by
\begin{eqnarray}\label{eqdensprob}
\rho=i(\psi^{*}_{5} \psi_{1}-\psi^{*}_{1} \psi_{5})
\end{eqnarray}
which coincides from Eq.(\ref{eqwavexpl}) with the standard expression 
$i(\psi^{*}_{5} \partial_{t}\psi_{5}-\partial_{t}\psi^{*}_{5} \psi_{5})$ obtained from the Klein-Gordon equation, except for an
unimportant normalization factor. When this is expressed in terms of $\psi_{\pm}$ one gets
\begin{eqnarray}\label{eqdensproba}
\rho=|\psi_{+}|^2-|\psi_{-}|^2
\end{eqnarray}
This quantity is not systematically positive. 

However in the non-relativistic regime, where the small component $\psi_{-}$ is negligible compared to 
the large component $\psi_{+}$, we recover a result proportional to 
$|\psi_{+}|^2=|\varphi_{+}|^2$, which is the expected expression for the density probability. 
One finds \cite{kemm} that this is $\psi_{+}$, rather than $\psi_{5}$, which is naturally coming in, but this is unimportant in the
non-relativistic regime. Moreover if we consider a particle, with positive energy, and small momentum, the large component
$\varphi_{+}$ is much larger than the small component $\varphi_{-}$, so its contribution is dominant in Eq.(\ref{eqdensproba})
and we still have $\rho>0$. But this conclusion remains valid even if we consider a bosonic particle with large velocity, which
is the physical case we have in mind. In this case we have important relativistic corrections to the Schr\"odinger equation,
described by Eq.(\ref{eqwavegen}). Nevertheless the large component is still dominant compared to the small one, and we
still have a positive density probability $\rho>0$ given by Eq.(\ref{eqdensproba}). Hence we have in this physical situation
an appropriate relativistic generalization of the Schr\"odinger equation, with a physically meaningful wave function describing
the presence probability of the particle. We note that, in this regime, we will need to consider $\varphi_{-}$
since, although its contribution is smaller than the one coming from $\varphi_{+}$, it is sizeable and we have to take it into
account.

The situation where $\psi_{-}$ is comparable or larger than $\psi_{+}$ corresponds physically to a different case, where the
negative energy states have to be interpreted as corresponding to the existence of anti-particles. In this case it is physically
normal that $\rho$ can take negative values, since it corresponds to the particle number minus the antiparticle number,
as it is found when anti-particles are properly introduced in the formalism by going to second quantization. In this process,
as it is well-known, the minus sign is unchanged since we deal with bosons and the coming-in operators are commuting.
This is in contrast with the situation in the Dirac equation, where the relevant operators satisfy anticommutation rules.
As a result a sign change appears when the second quantized expression of the particle number is written. As a consequence
the antiparticle number appears also with a minus sign in the particle number, as it is physically expected since the annihilation
of a particle with an antiparticle gives no particle at all.

In conclusion we have shown that, for a scalar boson, in the relativistic regime, one has to introduce a two-component wave
function, linked physically to the existence of positive and negative frequency eigenstates. This is in complete analogy with
the four-component wave function arising in the Dirac equation for the spin $1/2$ fermion. This two-component wave function
satisfies a differential equation Eq.(\ref{eqwavexplpm})
which is first order in time, again in full analogy with Dirac equation. When one goes to the regime
of velocities smaller than light velocity, one gets a dominant large component which satisfies a Schr\"odinger equation with
appropriate relativistic corrections. In this regime one has a positive probability density, obtained from the wave function, as
it is expected physically. In the limit of small velocities, one recovers in a natural way the standard Schr\"odinger equation.


\begin{references}
\bibitem{inter} Note that here we do not take interactions into account. However interactions with the electromagnetic field could
be introduced by making use of the standard minimal coupling, as it has been done by Kemmer \cite{kemm}.
\bibitem{expansion} This is naturally in agreeement with the expansion which can be obtained from the Klein-Gordon equation,
since from Eq.(\ref{eqklgord}) $\psi_{+}$ satisfies this equation.
\end{references}
\end{document}